\documentclass[aps,prl,twocolumn,showpacs,citeautoscript,reprint,longbibliography,superscriptaddress]{revtex4-1}
\usepackage{graphicx}
\usepackage{bm}
\usepackage{natbib}
\usepackage{amsmath}
\usepackage{xfrac}
\usepackage{nicefrac}
\usepackage[colorlinks=true, urlcolor=blue, linkcolor=blue, citecolor=blue, pdfborder={0 0 0}]{hyperref}
\usepackage[caption=false]{subfig}
\usepackage{cleveref}
\usepackage{dcolumn}
\usepackage{ulem}

\begin{document}

\title{Intrinsic mechanism for anisotropic magnetoresistance and experimental confirmation in Co$_x$Fe$_{1-x}$ single-crystal films}

\author{F. L. Zeng}
\altaffiliation{These authors contributed equally.}
\affiliation{Department of Physics, State Key Laboratory of Surface Physics, Fudan University, Shanghai 200433, China}
\author{Z. Y. Ren}
\altaffiliation{These authors contributed equally.}
\affiliation{School of Materials Science and Engineering, University of Science and Technology Beijing, Beijing 100083, China}
\affiliation{Center for Advanced Quantum Studies and Department of Physics, Beijing Normal University, Beijing 100875, China}
\author{Y. Li}
\affiliation{Department of Physics, Oakland University, Rochester MI 48309, USA}
\affiliation{Materials Science Division, Argonne National Laboratory, Lemont IL 60439, USA}
\author{J. Y. Zeng}
\affiliation{Department of Physics, State Key Laboratory of Surface Physics, Fudan University, Shanghai 200433, China}
\author{M. W. Jia}
\affiliation{Department of Physics, State Key Laboratory of Surface Physics, Fudan University, Shanghai 200433, China}
\author{J. Miao}
\affiliation{School of Materials Science and Engineering, University of Science and Technology Beijing, Beijing 100083, China}
\author{A. Hoffmann}
\affiliation{Materials Science Division, Argonne National Laboratory, Lemont IL 60439, USA}
\author{W. Zhang}
\affiliation{Department of Physics, Oakland University, Rochester MI 48309, USA}
\affiliation{Materials Science Division, Argonne National Laboratory, Lemont IL 60439, USA}
\author{Y. Z. Wu}
\email{wuyizheng@fudan.edu.cn}
\affiliation{Department of Physics, State Key Laboratory of Surface Physics, Fudan University, Shanghai 200433, China}
\affiliation{Collaborative Innovation Center of Advanced Microstructures, Nanjing 210093, China}
\author{Z. Yuan}
\email{zyuan@bnu.edu.cn}
\affiliation{Center for Advanced Quantum Studies and Department of Physics, Beijing Normal University, Beijing 100875, China}
\date{\today}

\begin{abstract}
Using first-principles transport calculations, we predict that the anisotropic magnetoresistance (AMR) of single-crystal Co$_x$Fe$_{1-x}$ alloys is strongly dependent on the current orientation and alloy concentration. An intrinsic mechanism for AMR is found to arise from the band crossing due to magnetization-dependent symmetry protection. These special $k$-points can be shifted towards or away from the Fermi energy by varying the alloy composition and hence the exchange splitting, thus allowing AMR tunability. The prediction is confirmed by delicate transport measurements, which further reveal a reciprocal relationship of the longitudinal and transverse resistivities along different crystal axes.
\end{abstract}

\maketitle

The spin-dependent transport properties of magnetic materials are the basis of spintronics devices used, for example, for magnetic sensing and data storage \cite{Chappert2007}. The electrical conductance of a magnetic device usually depends on its magnetization configuration, resulting in so-called magnetoresistance (MR) effects. Among these effects, AMR \cite{Doring1938,McGuire1975,Berger1963,Campbell1970} is fundamental for magnetic materials. It describes the dependence of the longitudinal electrical resistivity on the magnetization direction relative to the electric current in ferromagnetic materials.

AMR arises from the relativistic spin-orbit interaction (SOI), which couples the orbital motion of electrons with their spin angular momentum. The SOI leads to other spin transport phenomena, such as the anomalous Hall effect \cite{Nagaosa2010} and spin Hall effect \cite{Sinova2015}, whose microscopic mechanisms have been extensively studied experimentally and theoretically. The extrinsic contributions due to impurity scattering, including skew scattering and side jump, have been identified, as has the intrinsic mechanism that results from the Berry curvature \cite{Xiao2010} of the energy bands. In contrast, the microscopic understanding of AMR is still unsatisfactory after a long history of study, especially in single-crystal materials \cite{Bason2009,Limmer2006,Limmer2008,Naftalis2011,Ding2013,Tondra1993,VanGorkom2001,Hupfauer2015,Xiao20151,Xiao20152}. Phenomenologically, AMR can be described by a conductivity tensor, which is a function of the magnetization and current directions with respect to the crystallographic axes \cite{Doring1938,McGuire1975}. Alternatively, a two-current conduction model can be used to understand AMR, in which experimental values are usually needed to determine spin mixing parameters \cite{Campbell1970}.

Recently, many SOI-driven MRs have been discovered, including spin-Hall \cite{Nakayama2013}, Rashba \cite{Nakayama2016}, and spin-orbital MRs \cite{Zhou2018}, which also result in renewed interest in AMR, as it is a basic SOI-induced MR. Based on impurity scattering, some microscopic mechanisms for AMR have been identified, where the free-electron-like conduction bands were usually applied \cite{Trushin2009}. Nevertheless, the intrinsic band-structure effect on AMR that is fundamental in physics and applicable to pure ferromagnetic metals is not yet clear. The lack of a comprehensive understanding of AMR further hampers its manipulation and application in spintronics devices.

In this Letter, we take a single-crystalline Co$_{x}$Fe$_{1-x}$ alloy as an example and perform a joint experimental and theoretical study of its AMR effect. The CoFe alloy simultaneously has a large magnetization and very low damping \cite{Schoen2016,Lee2017} with strong anisotropy \cite{Li2019}, making it already an important material in industry. The calculated AMR exhibits a strong dependence on the current direction, and its amplitude is larger in the alloy regime than in the pure-metal limits. Detailed analysis reveals that the special $k$-points near Fermi energy play an essential role, where energy bands form crossing and anticrossing depending on the magnetization direction. This suggests an ``intrinsic'' mechanism for AMR arising from the band structure in addition to the ``extrinsic'' mechanisms based on the impurity-scattering picture. The predicted AMR properties are quantitatively confirmed by our transport experiments. A reciprocal relationship of the longitudinal and transverse resistivity is obtained along the $\langle110\rangle$ and $\langle100\rangle$ crystal axes.


\begin{figure}[ht]
\includegraphics[width=\linewidth]{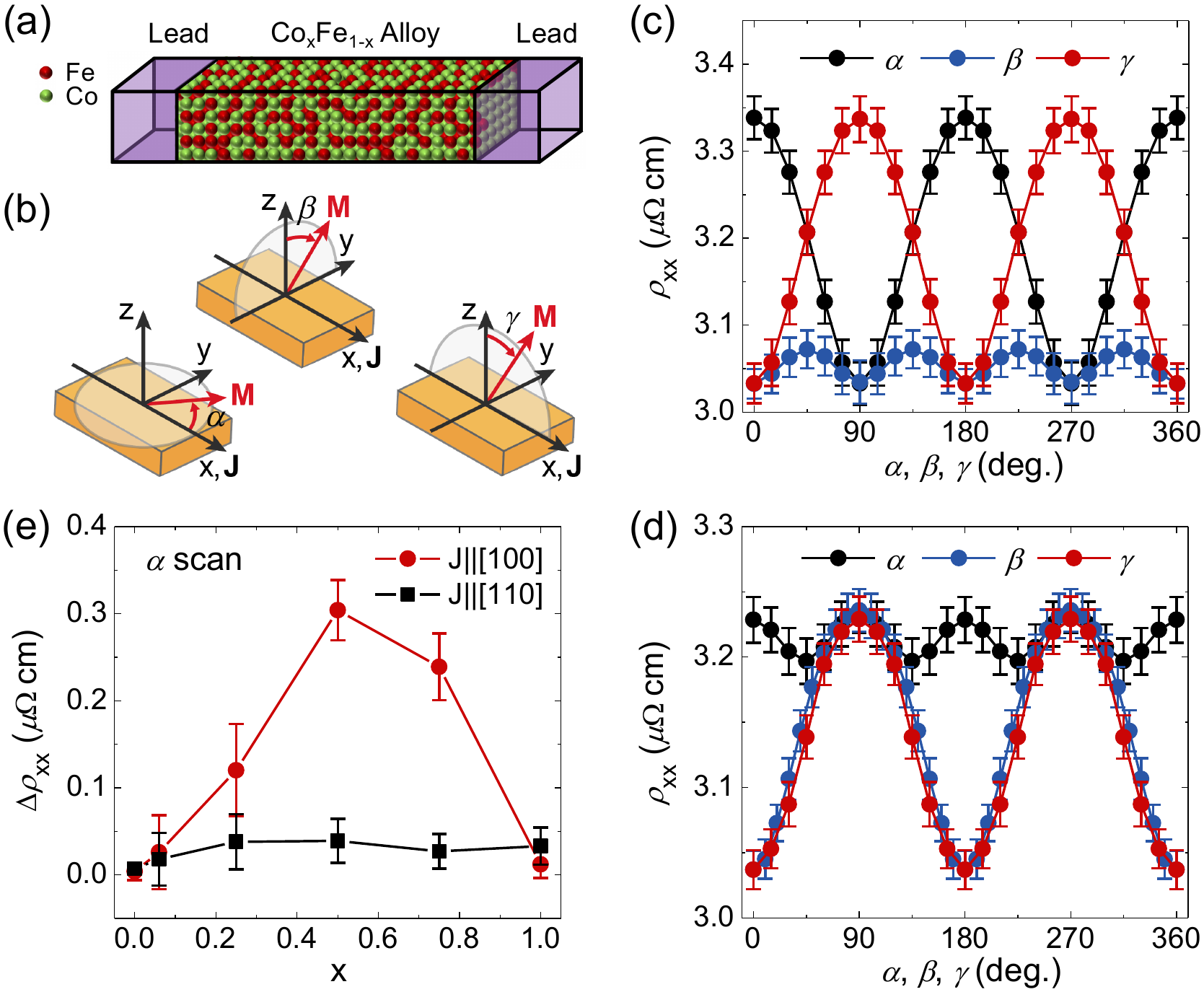}
\caption{(a) Sketch of the transport geometry in the calculation. (b) Definition of angles $\alpha$, $\beta$ and $\gamma$ in Cartesian coordinates. The electric current $\mathbf J$ is always along the $x$ axis. Calculated $\rho_{xx}$ of Co$_{0.5}$Fe$_{0.5}$ as a function of $\alpha$, $\beta$ and $\gamma$ for $\mathbf J\parallel[100]$ (c) and $\mathbf J\parallel[110]$ (d). (e) Largest variation in $\Delta\rho_{xx}$ of the Co$_{x}$Fe$_{1-x}$ alloy when varying $\alpha$ as a function of Co concentration $x$.}
\label{fig1}
\end{figure}
{\color{red}\it AMR from first principles.---}The resistivity $\rho_{xx}$ of single-crystal Co$_x$Fe$_{1-x}$ alloy is calculated using the first-principles Landauer-B{\"u}ttiker formalism including the SOI \cite{Starikov2018,SM}. For $\mathbf J$ along $[100]$, we use a Cartesian coordinate system with $x\parallel[100]$, $y\parallel[010]$ and $z\parallel[001]$. Then, the calculated $\rho_{xx}$ of Co$_{0.5}$Fe$_{0.5}$ is plotted in Fig.~\ref{fig1}(c) as a function of the $\mathbf M$ direction, where the angles $\alpha$, $\beta$ and $\gamma$ are explicitly defined in Fig.~\ref{fig1}(b). As $\alpha$ or $\gamma$ varies, $\rho_{xx}$ exhibits a two-fold symmetry, and the maximum (minimum) of $\rho_{xx}$ occurs for $\mathbf J\parallel\mathbf M$ ($\mathbf J\perp\mathbf M$). A much weaker four-fold symmetry is found with varying $\beta$. The resistivity is shown in Fig.~\ref{fig1}(d) for $\mathbf J\parallel x\parallel[110]$, $y\parallel[\bar{1}10]$, and $z\parallel[001]$, where $\rho_{xx}(\alpha)$ shows a weak four-fold symmetry, in sharp contrast to the case of $\mathbf J\parallel[100]$. The variations in $\rho_{xx}$ are larger with rotation of $\beta$ and $\gamma$, and both exhibit a two-fold symmetry. The calculated resistivity for $\mathbf J\parallel[110]$ shows an interesting relationship of $\rho_{x}\approx\rho_{y}>\rho_{z}$, which is in sharp contrast to the ordinary AMR relationship $\rho_{x}>\rho_{y}=\rho_{z}$ and has never been reported for any MRs.

We plot the largest variation in the resistivity $\Delta\rho_{xx}\equiv\max[\rho_{xx}(\alpha)]-\min[\rho_{xx}(\alpha)]$ in Fig.~\ref{fig1}(e) as a function of Co concentration $x$. Here, a significant difference is seen for $\mathbf J\parallel[110]$ and $\mathbf J\parallel[100]$: $\Delta\rho^{[110]}_{xx}$ is at most 0.05~$\mu\Omega\,\mathrm{cm}$ for all concentrations, while $\Delta\rho^{[100]}_{xx}$ is as large as 0.3~$\mu\Omega\,\mathrm{cm}$ at $x=0.5$. In addition, this giant current-orientation-dependent AMR is found to be more pronounced in alloys than in pure metals. This is counterintuitive because one would expect that the random arrangement of Co and Fe atoms in Co$_x$Fe$_{1-x}$ alloys would lower the crystalline symmetry of pure metals.

\begin{figure}[t]
\includegraphics[width=1.0\linewidth]{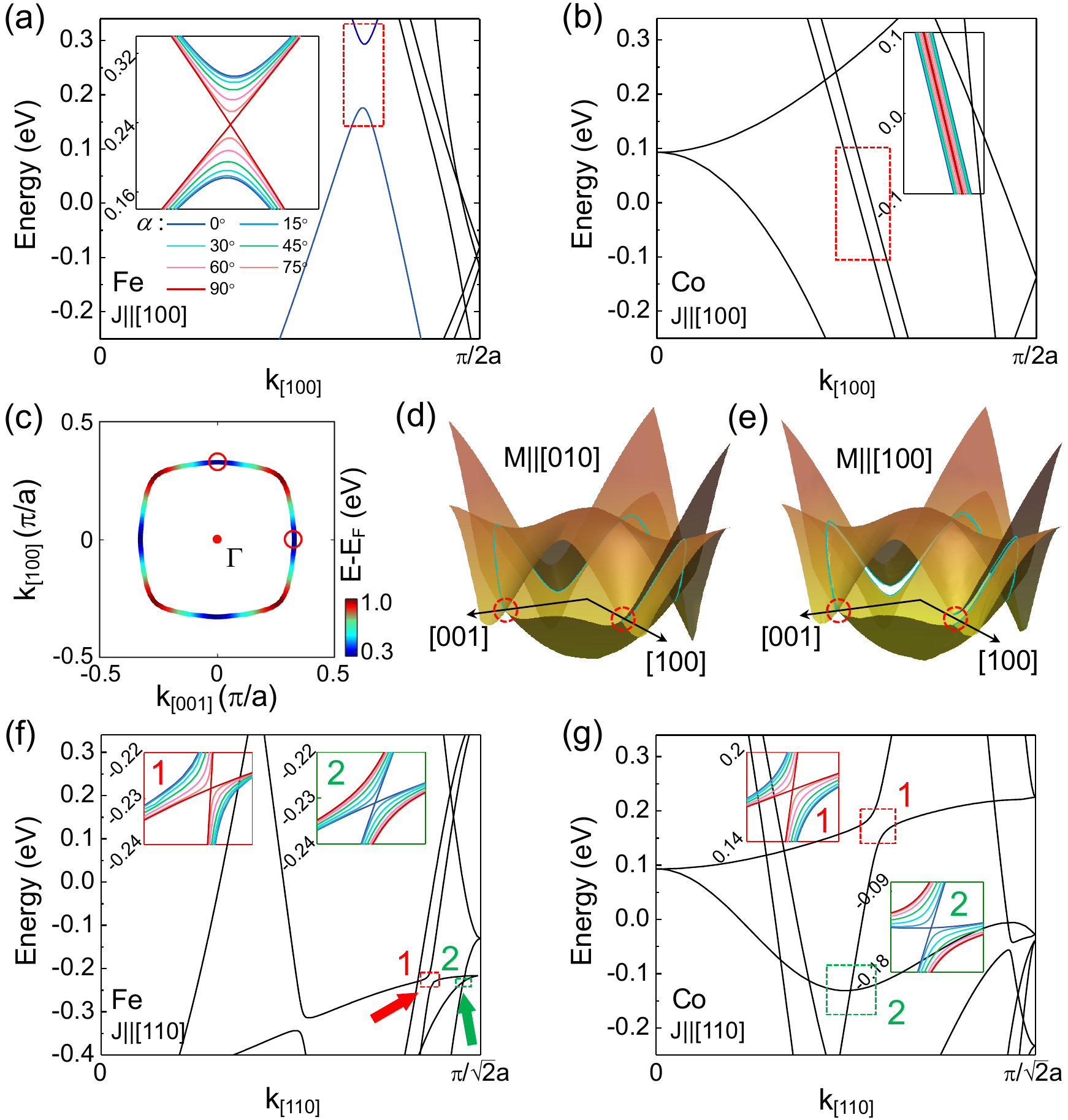}
\caption{Calculated band structure along $[100]$ using the effective potentials of Fe (a) and Co (b) in Co$_{0.5}$Fe$_{0.5}$. (c) Nodal line in the (010) plane, to which the band crossing in (a) at $\alpha=90^\circ$ belongs. Dispersion of the two bands forming the nodal line at $\alpha=90^\circ$ (d) and $\alpha=0^\circ$ (e). The nodal line disappears in (e), and an anticrossing band gap appears except for two special $k$-points along [001], as highlighted by red circles. Calculated band structure along $[110]$ using the effective potentials of Fe (f) and Co (g) in Co$_{0.5}$Fe$_{0.5}$. In (a), (b), (f) and (g), only the bands marked by frames change with $\alpha$, which are shown in the insets.}
\label{fig2}
\end{figure}
{\color{red}\it The intrinsic mechanism for AMR.---}To understand the calculated AMR and unravel its microscopic nature, we focus on the electronic structure of bcc Co$_x$Fe$_{1-x}$ alloys. Applying the coherent potential approximation, we self-consistently compute auxiliary potentials for Co and Fe in Co$_x$Fe$_{1-x}$ alloys \cite{Turek1997}, and these effective potentials are randomly distributed in the transport calculations. It is instructive to place the ``effective Fe'' potential on a perfect bcc lattice and non-self-consistently calculate the band structure \cite{Starikov2018}. Then, we perform the same calculation for bcc Co using the ``effective Co'' potential. These calculated band structures reflect the averaged electronic properties of Fe and Co atoms, whereas the bands are smeared in alloys due to the random arrangement of Fe and Co atoms.

The bands of Fe in Co$_{0.5}$Fe$_{0.5}$ along $[100]$ for $\alpha=0^\circ$ are plotted in Fig.~\ref{fig2}(a), where a gap of $\sim0.1~$eV appears near $E_F$, as highlighted by the red frame. When the magnetization rotates from $\alpha=0^\circ$ to $90^\circ$, this gap shrinks until reaching a crossing point; see the inset. All the other bands along $[100]$ near $E_F$ do not depend on $\alpha$. The crossing point at $\alpha=90^\circ$ belongs to the nodal line located in the $(010)$ plane with $\mathbf M\parallel [010]$, as shown in Fig.~\ref{fig2}(c) and (d). This nodal line forms a closed ring around the $\Gamma$ point and has the lowest energy close to $E_F$ with the $\mathbf k$ vector along $\langle100\rangle$, indicating its largest influence on the electronic transport for the current along $\langle100\rangle$. Since the magnetization breaks the time-reversal symmetry, this nodal line is protected by the mirror symmetry \cite{Yang2018} about a crystalline plane perpendicular to $\mathbf M$. By rotating $\mathbf M$, the previous mirror symmetry is broken, and the nodal line disappears; see Fig.~\ref{fig2}(e). Instead, an anticrossing gap appears except for at two special $k$-points along $[001]$. When $\mathbf M$ is rotated to $[100]$, another nodal line forms in the $(100)$ plane. Therefore, the closing and opening of the band gap in Fig.~\ref{fig2}(a) when rotating $\mathbf M$ can be understood based on the required symmetry of the nodal line.

Around the crossing points, the two bands have different topological characteristics and do not interact with each other. Thus, the interband scattering has a relatively low probability. When the bands interact and form an anticrossing, the interband scattering rate increases such that the resistivity becomes larger \cite{Syzranov2017,Sbierski2014,DasSarma2011}. From the quasi-particle point of view, the long effective wavelengths and small effective masses of the topological states have greater probabilities of surviving the back-scattering caused by disorder than other non-topological Bloch states \cite{Jiang2015,Jiang2016}. Therefore, $\rho_{xx}$ in Fig.~\ref{fig1}(c) monotonically decreases with increasing $\alpha$ and reaches the minimum at $\alpha=90^\circ$. If we artificially shift $E_F$ 0.2~eV upwards such that it approaches the crossing point, then the calculated $\Delta\rho_{xx}/\rho_{xx}$ increases by 12\%. This numerical test confirms the correlation of the AMR and the $\mathbf M$-dependent band crossing \cite{SM}. Recently, the effect of the band topology on spin-dependent transport has been discussed in antiferromagnetic spintronics \cite{Smejkal2018,Bodnar2018}.

Along $[110]$, the Fe bands at two special $k$-points depend on the magnetization direction, which are marked by the frames with labels 1 and 2 in Fig.~\ref{fig2}(f). At $k$-point 1, a gap appears at $\alpha=0^\circ$ but closes at $\alpha=90^\circ$. Conversely, the opposite $\alpha$ dependence occurs for the gap at $k$-point 2. The Co bands along $[110]$ in Fig.~\ref{fig2}(g) also have opposite $\alpha$ dependences at two $k$-points near $E_F$. The competing effects at these $k$ pairs result in a non-monotonic variation in the resistivity for $\mathbf J\parallel[110]$ when $\alpha$ increases from $0^\circ$ to $90^\circ$. Thus, the ordinary two-fold AMR is suppressed, and $\Delta\rho^{[110]}_{xx}$ is much smaller than $\Delta\rho^{[100]}_{xx}$. Disorder scattering in alloys breaks the momentum conservation, resulting in effective band broadening; therefore, the special $k$-points can affect the electrical resistivity, although they are not located precisely at the Fermi energy. Such band analysis is applicable in explanation of all the angular dependence shown in Fig.~\ref{fig1}~\cite{SM}.

The crossing points can be shifted up or down by varying the Co concentration $x$. When the crossing points move closer to $E_F$, their contribution to $\rho_{xx}$ increases. For example, the band crossing in the Fe band along $[100]$ gradually shifts down towards $E_F$ with increasing $x$ \cite{SM}; therefore, the AMR along $[100]$ in Fig.~\ref{fig1}(e) is more pronounced in the alloys than that in pure Fe. The calculated Co bands do not show significant variation for $x$ up to 1, and the pure Co has a very small AMR, indicating that the band changes without gap closing and opening have little effect on the AMR.

We shall distinguish underlying physics for the quantities depending on the SOI-mediated band structure. The anomalous Hall effect results from Berry curvature at anticrossing bands, which contributes to the anomalous velocity \cite{Xiao2010,Nagaosa2010}. For Gilbert damping \cite{Gilmore2010,Ebert2011,Yuan2014,Starikov2018} and magnetic inertia \cite{Thonig2017,Nikolic2019}, the SOI lifts bands across Fermi energy back and forth with rotating magnetization as described by the breathing Fermi surface model \cite{Kambersky1976}. The intrinsic mechanism for AMR here comes from the symmetry protected topological states, which determines the dependence of longitudinal transport on crystal and magnetization directions.

\begin{figure}[t]
\includegraphics[width=1.0\linewidth]{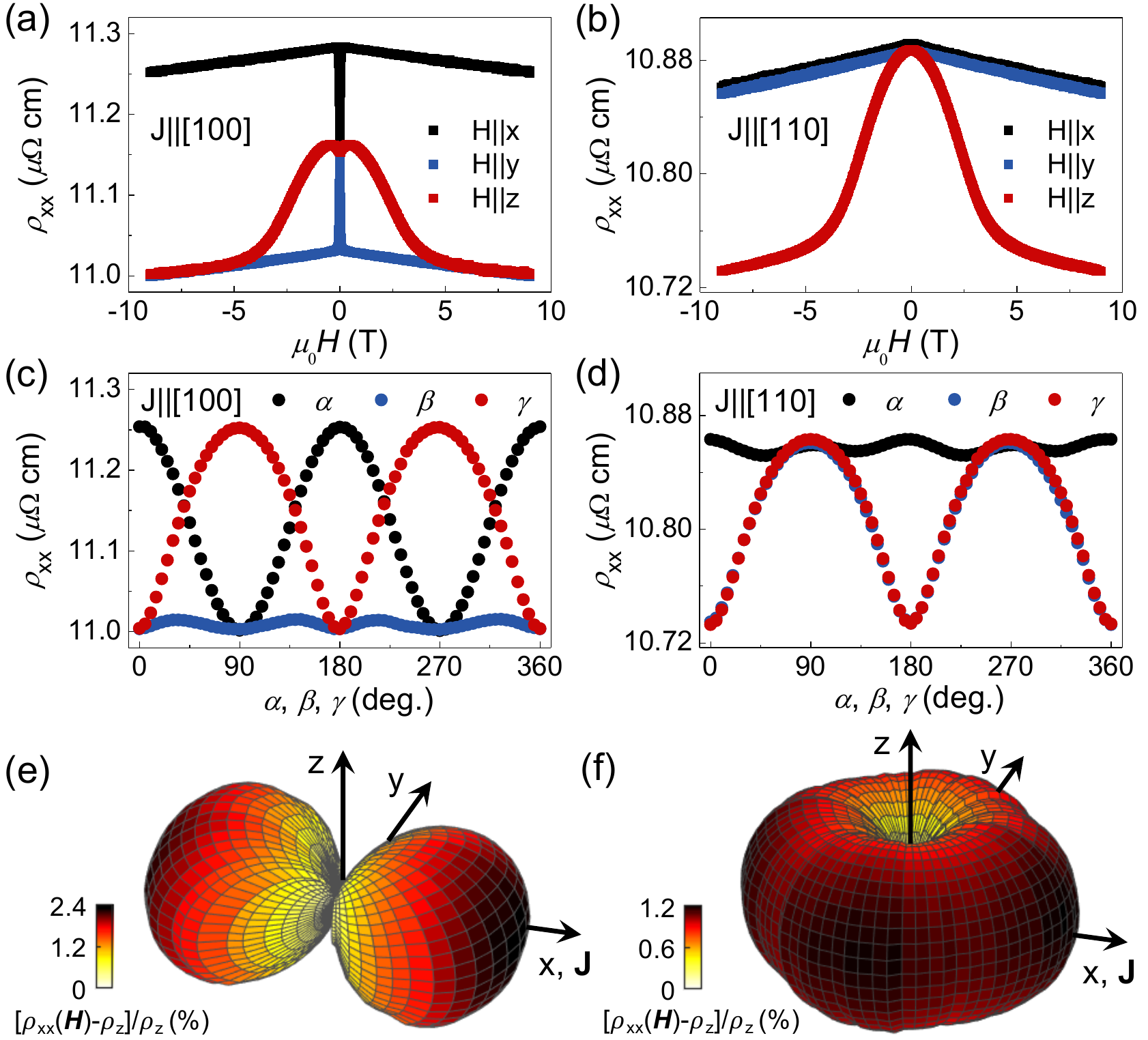}
\caption{Experimental resistivity along $[100]$ (a) and $[110]$ (b) of a 10-nm-thick Co$_{0.5}$Fe$_{0.5}$ sample measured under a magnetic field along principal axes $x$, $y$ and $z$. The $x$ axis is always defined as being along the current direction. Measured resistivity with rotating magnetization in the $xy$, $yz$ and $xz$ planes under the application of a 9-Tesla field with $\mathbf J\parallel[100]$ (c) and $\mathbf J\parallel[110]$ (d). The angles $\alpha$, $\beta$ and $\gamma$ are defined in Fig.~\ref{fig1}(b). Three-dimensional plots of the MR for the current along $[100]$ (e) and $[110]$ (f).}
\label{fig3}
\end{figure}
{\color{red}\it Experimental measurements.---}To verify the theoretical calculation and analysis, we performed AMR measurements on single-crystalline Co$_{x}$Fe$_{1-x}$ film deposited on MgO(001) substrates by molecular beam epitaxy \cite{SM}. Figure~\ref{fig3}(a) and (b) shows the measured resistivity of the Co$_{0.5}$Fe$_{0.5}$ film for $\mathbf J\parallel[100]$ and $\mathbf J\parallel[110]$, respectively, as a function of the external magnetic field, which is applied along the three principal axes. At a sufficiently large field, the measured $\rho_{xx}$ linearly decreases with increasing field $H$, and this decrease can be attributed to the field-induced suppression of electron-magnon scattering \cite{Raquet2002}.

We then performed transport measurements along $[100]$ and $[110]$ by rotating the applied 9-Tesla field in the $xy$, $yz$ and $xz$ planes, separately. As shown in Fig.~\ref{fig3}(c), a strong two-fold symmetry of $\rho_{xx}$ is seen as a function of $\alpha$ and $\gamma$ for $\mathbf J\parallel[100]$, while a weak four-fold symmetry appears for the $\beta$ scan. In contrast, for $\mathbf J\parallel[110]$, as shown in Fig.~\ref{fig3}(d), a strong two-fold symmetry occurs in the $\beta$ and $\gamma$ scans, and a weak four-fold symmetry is obtained when varying $\alpha$.

\begin{figure}[t]
\includegraphics[width=1.0\linewidth]{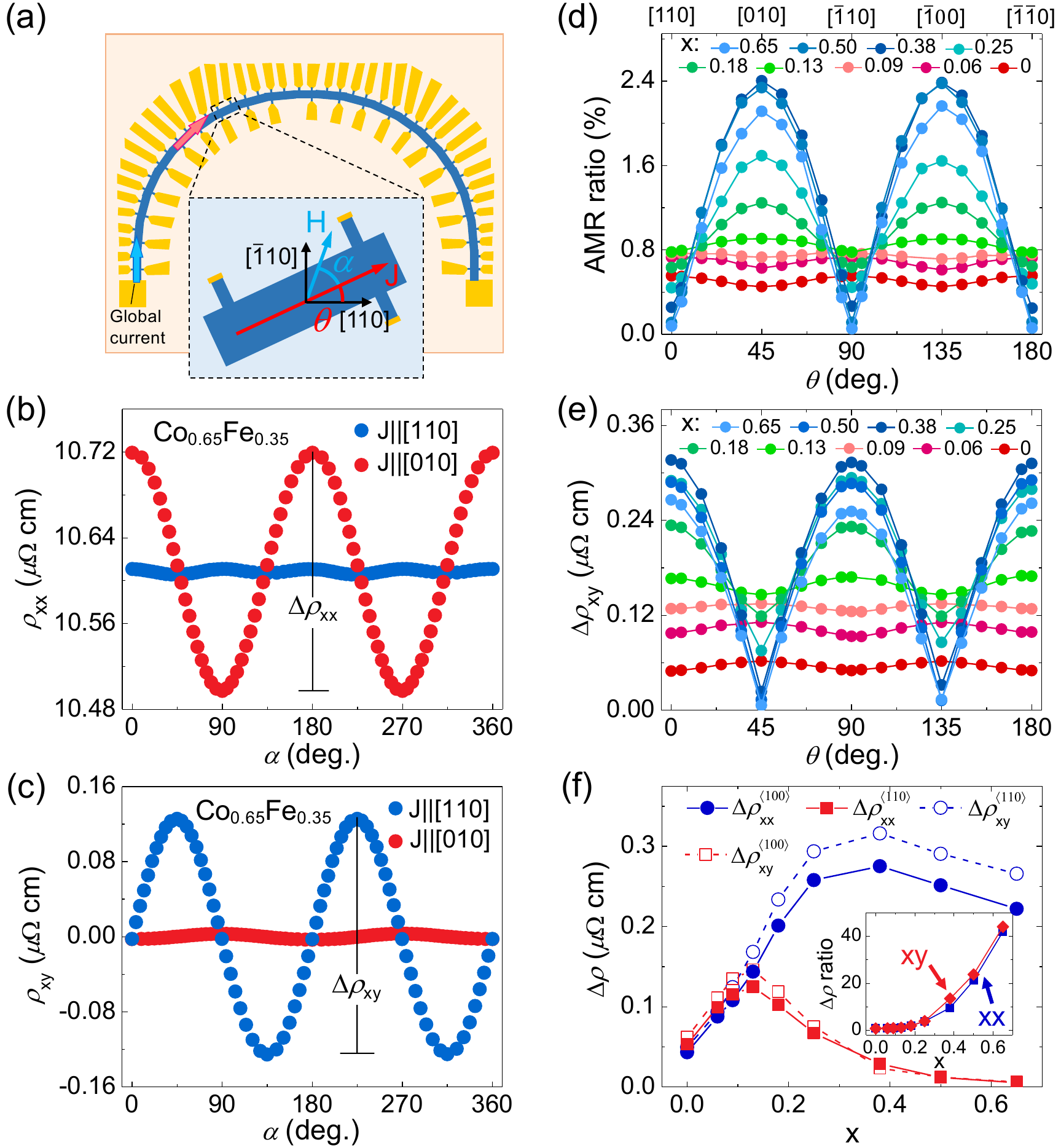}
\caption{(a) Schematic illustration of the device. The angle $\theta$ defines the current orientation direction with respect to Co$_x$Fe$_{1-x}[110]$. (b) $\rho_{xx}$ and (c) $\rho_{xy}$ measured on a 10-nm-thick Co$_{0.65}$Fe$_{0.35}$ alloy as a function of magnetic field direction for current along [110] and [010]. Current-orientation-dependent AMR ratio (d) and $\Delta\rho_{xy}$ (e) for different Co concentrations $x$. (f) The current-orientation-dependent resistivity changes as a function of $x$. Inset: concentration-dependent ratios $\Delta\rho_{xx}^{\langle100\rangle}/\Delta\rho_{xx}^{\langle110\rangle}$ (blue) and $\Delta\rho_{xy}^{\langle110\rangle}/\Delta\rho_{xy}^{\langle100\rangle}$ (red).}
\label{fig4}
\end{figure}
We then measured the MR under the field along an arbitrary direction by rotating the sample. The three-dimensional plots of the measured relative AMR, $[\rho_{xx}(\mathbf H)-\rho_z]/\rho_z$ with $\rho_z\equiv\rho_{xx}(H\hat z)$, are shown in Fig.~\ref{fig3}(e) and (f). The angular dependence shows a dumbbell shape for $\mathbf J\parallel[100]$, which is expected for the AMR in most ferromagnetic materials. Nevertheless, the resistivity for $\mathbf J\parallel[110]$ exhibits a donut shape that has never been previously reported in literature and confirms the calculated relationship $\rho_{x}\approx\rho_{y}>\rho_{z}$. Note that the different angular-dependent AMRs in Fig.~\ref{fig3} are obtained with the same sample, indicating that the current-orientation effect arises from the electronic structure due to the anisotropic crystal field. The angular dependences of the resistivity measured in the experiment fully agree with the calculated results in Fig.~\ref{fig1}. Despite of the significant variation of $\rho_{xx}$ with temperature and sample thickness, the quantitative agreement in the experimental and calculated $\Delta\rho_{xx}$ further confirms the predicted intrinsic nature \cite{SM}. At very small thickness, the emergence of interfacial spin-orbit field \cite{Hupfauer2015} may have additional effect on AMR, which is beyond the scope of this work.

We further developed our experiment to simultaneously measure the longitudinal and transverse resistivity under an arbitrary current orientation, as schematically shown in Fig.~\ref{fig4}(a). The single-crystal Co$_x$Fe$_{1-x}$ films deposited on MgO(001) substrates were patterned into $300~\mu\mathrm{m}\times 100~\mu\mathrm{m}$ Hall bars with continuously varying current directions. For $x=0.65$, the experimental $\rho_{xx}$ and $\rho_{xy}$ are plotted in Fig.~\ref{fig4}(b) and (c), which exhibit a reciprocal relationship. For $\mathbf J\parallel[010]$, the largest variation in longitudinal resistivity $\Delta\rho_{xx}$ is large when rotating $\mathbf H$, and the corresponding variation in transverse resistivity $\Delta\rho_{xy}$ is small. The opposite relationship of their amplitudes is found for $\mathbf J\parallel[110]$.

The measured AMR ratios defined by $\Delta\rho_{xx}/\min(\rho_{xx})$ and $\Delta\rho_{xy}$ with different Co concentrations are plotted in Fig.~\ref{fig4}(d) and (e), respectively, both as a function of current orientation. Here, again, a reciprocal relationship between $\Delta\rho_{xx}$ and $\Delta\rho_{xy}$ is unambiguously demonstrated: at a given $\theta$ where $\Delta\rho_{xy}$ has its maximum amplitude, $\Delta\rho_{xx}$ has its minimum value, and vice versa. This reciprocal relationship is analytically reproduced \cite{SM} by the phenomenological expansion based on symmetry \cite{Birss1964}.

The AMR ratio exhibits a strong Co concentration dependence. At small $x$, the AMR ratio is nearly independent of the current direction $\theta$, but for $x>0.25$, a giant difference between the maximum at $\theta=45^\circ$ and $135^\circ$ and the minimum at $\theta=0^\circ$ and $180^\circ$ is observed. To quantitatively elucidate the concentration dependence, we plot $\Delta\rho_{xx}$ and $\Delta\rho_{xy}$ as a function of $x$ in Fig.~\ref{fig4}(f). For the current along $\langle100\rangle$, $\Delta\rho_{xx}$ increases with increasing $x$ up to 0.38 and slightly decreases for larger $x$. In contrast, $\Delta\rho_{xx}^{\langle110\rangle}$ has a relatively small amplitude. This current-orientation dependence and concentration dependence of $\Delta\rho_{xx}$ are both in very good agreement with the theoretical calculation in Fig.~\ref{fig1}(e) except for the small bump at $x=0.13$ in the red curves. This bump may be attributed to the inhomogeneity in the alloy samples at small $x$ while homogeneous mixing is assumed in calculation. Moreover, the ratio between the experimental $\Delta\rho_{xx}^{\langle100\rangle}$ and $\Delta\rho_{xx}^{\langle110\rangle}$ monotonically increases with $x$ and is as large as 42 at $x=0.65$, as shown in the inset of Fig.~\ref{fig4}(f). Such a large anisotropic dependence on the current orientation has never been previously reported for the AMR. Following the reciprocal relation, $\Delta\rho_{xy}^{\langle110\rangle}/\Delta\rho_{xy}^{\langle100\rangle}$ has the same dependence on $x$.

{\color{red}\it Conclusions.---}We have calculated the AMR in single-crystal Co$_x$Fe$_{1-x}$ alloys using a first-principles transport formalism. Our band structure analysis unambiguously identifies an intrinsic contribution to the AMR: the energy band crossing depends on the magnetization direction with spin-orbit coupling. The predicted properties of the AMR in Co$_x$Fe$_{1-x}$ alloys, including its dependence on the current orientation and alloy concentration, are well confirmed by our transport experiments on single-crystal samples. The simultaneously measured longitudinal and transverse resistivities in the experiment exhibit a reciprocal relationship along high-symmetry crystal axes, which is reproduced by a phenomenological model.

\acknowledgements
The work at Fudan University was supported by the National Key Research and Development Program of China (Grant No. 2016YFA0300703), National Natural Science Foundation of China (Grant No. 11974079 and No. 11734006), and the Program of Shanghai Academic Research Leader (No. 17XD1400400). The work at Beijing Normal University was supported by the National Natural Science Foundation of China (Grant No. 61774018 and No. 11734004), the Recruitment Program of Global Youth Experts, and the Fundamental Research Funds for the Central Universities (Grant No. 2018EYT03). The work at Argonne was supported by the U.S. Department of Energy (DOE), Office of Science, Materials Science and Engineering Division. The work at OU was supported by the U.S. National Science Foundation under Grant No. DMR-1808892.

\end{document}